\documentclass[aps,a4paper,twocolumn,nofootinbib,superscriptaddress]{revtex4}
\usepackage{amsmath,amssymb,color,dsfont}
\usepackage[font=footnotesize,caption=false]{subfig}

\usepackage{pdfpages} 

\newcommand{\nbar}{\ensuremath{\bar{n}}}

\newcommand{\bra}[1]{\ensuremath{\left\langle #1 \right|}}
\newcommand{\ket}[1]{\ensuremath{\left| #1 \right\rangle}}
\newcommand{\bras}[1]{\ensuremath{\langle #1 |}}
\newcommand{\kets}[1]{\ensuremath{| #1 \rangle}}

\newcommand{\PL}{\ensuremath{P_\textrm{L}}}

\newcommand{\etal}{\emph{et al}.}

\newcommand{\PRL}[3]{Phys.\@ Rev. Lett.~\textbf{#1}, #2 (#3)}
\newcommand{\PRA}[3]{Phys. Rev. A~\textbf{#1}, #2 (#3)}
\newcommand{\PRAR}[3]{Phys. Rev. A~\textbf{#1}, #2(R) (#3)} 

\newcommand{\PRX}[3]{Phys. Rev. X~\textbf{#1}, #2 (#3)}
\newcommand{\RMP}[3]{Rev. Mod. Phys.~\textbf{#1}, #2 (#3)}

\newcommand{\Nature}[3]{Nature~\textbf{#1}, #2 (#3)}
\newcommand{\NatComm}[3]{Nat. Commun.~\textbf{#1}, #2 (#3)}

\newcommand{\NatPhot}[3]{Nature Photon.~\textbf{#1}, #2 (#3)}

\newcommand{\SREP}[3]{Sci. Rep.~\textbf{#1}, #2 (#3)}

\newcommand{\Annalen}[3]{Ann. Phys. (Berlin)~\textbf{#1}, #2 (#3)}
\newcommand{\APL}[3]{Appl. Phys. Lett.~\textbf{#1}, #2 (#3)}

\newcommand{\JOptB}[3]{J. Opt. B~\textbf{#1}, #2 (#3)}
\newcommand{\JPhysB}[3]{J. Phys. B~\textbf{#1}, #2 (#3)}

\newcommand{\PNAS}[3]{Proc. Nat. Acad. Sci. USA~\textbf{#1}, #2 (#3)}
\newcommand{\RPG}[3]{Rep. Prog. Phys.~\textbf{#1}, #2 (#3)}
\newcommand{\Science}[3]{Science~\textbf{#1}, #2 (#3)}

\begin{document}

\title{
Non-classical state generation in macroscopic systems via hybrid discrete-continuous quantum measurements
}

\author{T. J. Milburn}
\email{thomas.milburn@ati.ac.at}
\affiliation{Institute of Atomic and Subatomic Physics, TU Wien, Stadionallee 2, 1020 Wien, Austria}

\author{M. S. Kim}
\affiliation{QOLS, Blackett Laboratory, Imperial College London, London SW7 2BW, United Kingdom}
\affiliation{Korea Institute of Advanced Study, Seoul, 130-722, Korea}

\author{M. R. Vanner} 
\email{michael.vanner@physics.ox.ac.uk}
\affiliation{School of Mathematics and Physics, The University of Queensland, Brisbane, Queensland 4072, Australia}
\affiliation{Clarendon Laboratory, Department of Physics, University of Oxford, OX1 3PU, United Kingdom}

\date{\today}

\begin{abstract}
Non-classical state generation is an important component throughout experimental quantum science for quantum information applications and probing the fundamentals of physics. Here, we investigate permutations of quantum non-demolition quadrature measurements and single quanta addition/subtraction to prepare quantum superposition states in bosonic systems. The performance of each permutation is quantified and compared using several different non-classicality criteria including Wigner negativity, non-classical depth, and optimal fidelity with a coherent state superposition. We also compare the performance of our protocol using squeezing instead of a quadrature measurement and find that the purification provided by the quadrature measurement can significantly increase the non-classicality generated. Our approach is ideally suited for implementation in light-matter systems such as quantum optomechanics and atomic spin ensembles, and offers considerable robustness to initial thermal occupation.
\end{abstract}

\maketitle

\section{Introduction}

Probing the foundations of quantum mechanics as well as developing a powerful suite of quantum-enhanced technologies are two current major themes of experimental quantum science. To these ends, generating and studying non-classicality in massive macroscopic systems allows for a greater understanding of the quantum-to-classical transition as well as the development of quantum information and metrology applications. Experimental efforts towards these goals are diversifying and gaining increasing interest with example physical platforms now including matter-wave interferometers~\cite{Juffmann2013}, atomic spin-ensembles~\cite{Hammerer2010}, superconducting circuits~\cite{Clarke2008, Devoret2013}, and cavity opto-mechanical systems~\cite{Meystre2013, Aspelmeyer2014}.

In quantum optics, one of the key tools used for quantum state preparation is quantum measurement. With this approach, a quantum state is conditionally generated by a measurement and the state's properties are created by a combination of Bayesian inference and quantum back-action. Quantum-measurement-based state preparation has been used extensively and to great success in purely optical experiments to: prepare squeezed states via quadrature back-action evading or quantum non-demolition (QND) measurement~\cite{Levenson1986, Porta1989}, generate quantum non-Guassian states by photon addition and subtraction~\cite{Wenger2004, Ourjoumtsev2006, Neergaard2006, Zavatta2007, Parigi2007, Kim2008}, and utilising the path entanglement from a quantum state on a beam-splitter to prepare non-classical states with linear measurement~\cite{Babichev2004, Ourjoumtsev2007}. Quantum measurement offers considerable versatility to prepare a large range of different quantum states and, e.g., excellent approximations to the coherent state superposition, or cat state, can be generated. This state is studied extensively throughout quantum optics and is of particular interest for quantum information~\cite{Jeong2002, Ralph2003} and quantum metrology~\cite{Munro2002} applications.

\begin{figure}[b]%
	\begin{center}
		\includegraphics[width=\columnwidth]{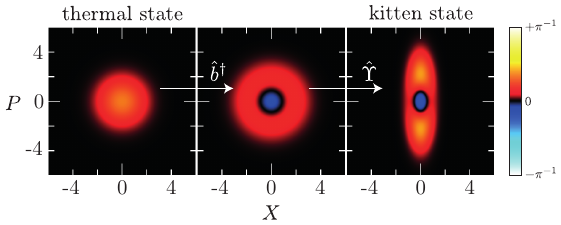}
		\caption{(Colour online.)  Example quantum state preparation protocol using single quanta addition followed by a quadrature measurement. The initial state is a low occupation thermal state (Gaussian Wigner function shown on the left). Single quanta addition, $\hat{b}^\dag$, creates a region of Wigner negativity (blue) at the centre of the distribution (shown in the centre).  Finally, we perform an $X$-quadrature measurement, $\hat{\Upsilon}$, selecting outcomes close to $X = 0$, which creates a `kitten' state---approximately a cat state (shown on the right).  This and other permutations of quantum measurements are studied here.}
		\label{fig:extraIntroductionFigure}
	\end{center}
\end{figure}

Building from the techniques developed in optics, quantum measurement is now being explored for non-classical state preparation of matter-based systems such as atomic spin ensembles and the motion of mechanical resonators. For atomic ensembles, spin squeezing by measurement is now an established technique~\cite{Kuzmich2000, Appel2009, Sewell2013} and very recently heralded single quanta addition has been employed to generate non-Gaussian states~\cite{Christensen2014} with significant Wigner negativity~\cite{McConnell2015}. In optomechanics, squeezing of a mechanical quadrature of motion can be achieved deterministically by using a parametric modulation, which has been used to demonstrate squeezing of thermal motion~\cite{Rugar1991}. Alternatively, squeezing can be achieved by a back-action evading measurement such as a stroboscopic~\cite{BraginskyBook} or two-toned measurement~\cite{Suh2014, Lecocq2015}. Pulsed interactions also allow for cooling and squeezing by measurement as well as quantum state tomography~\cite{Vanner2011, Vanner2013}, and can be used in sequence to generate a quantum optomechanical interface~\cite{Marek2010, Bennett2015, Rakhubovsky2016}. Though much of the current focus of optomechanics is on linear interactions and measurement, non-Guassian operations are now also being explored by conditioning on single photon counting. By analogy to heralded single photon generation using parametric down-conversion in optics, a cavity optomechanical scheme for heralded single phonon addition and subtraction by single photon detection was proposed in Ref.~\cite{VannerPRL2013}. There, the detection of a single photon shifted down (up) by the mechanical frequency heralds a phonon addition (subtraction) process. Single photon counting was used to generate a non-classical state of motion of the Terahertz frequency lattice vibration in a diamond crystal~\cite{Lee2012} and experimental progress in optomechanics using single photon detection has very recently been made~\cite{Cohen2015, Riedinger2015}. State generation by quantum measurement is also very useful for multi-mode quantum state engineering. Indeed, using a single photon detection event, entanglement between the vibrational states of two diamond crystals has been generated~\cite{Lee2011}, and techniques for mechanical $N00N$ state generation have been proposed~\cite{Ren2013, Flayac2014}, as well as schemes to entangle atomic ensembles with mechanical motion~\cite{Hammerer2009, Zhang2014}. In addition, a scheme has been recently proposed that utilises an optomechanical interaction with a non-classical optical field prepared via photon subtraction to conditionally generate non-classical mechanical states~\cite{Hoff2016}.

Unlike optical fields, massive bosonic systems are often incoherently excited due to their lower resonance frequency. For example, mechanical oscillators can undergo thermal Brownian motion, and spin ensembles may have imperfect spin polarization. Such impurity adds to the challenge of non-classical state engineering in these systems and has largely been neglected in the theoretical optics literature for non-classical state generation. Here, we analyze the use of QND quadrature measurements in combination with single quanta addition or subtraction for non-classical state generation in bosonic systems (see Fig.~\ref{fig:extraIntroductionFigure}). The use of measurement instead of parametric squeezing alleviates the challenge of initial state impurity and also allows the protocol to be applied in light-matter systems where QND quadrature measurements are more easily implemented. We analyze the states prepared by the different permutations of these operations and compare them to the states that can be generated using squeezing instead of a quadrature measurement. For a finite thermal occupation, it is a non-trivial problem to identify which sequence of these operations yields the greatest non-classicality. The use of a QND quadrature measurement for this type of protocol has not been previously considered and, as is detailed below, offers several advantages. We characterise the non-classicality generated by our scheme using three figures of merit: (i) Wigner negativity~\cite{Kenfack2004}; (ii) non-classical depth~\cite{Lee1991, Lutkenhaus1995}; and (iii) optimal fidelity with a cat state~\cite{Kim2005}. We describe our scheme primarily in the context of cavity optomechanics, however, our results are directly applicable to atomic spin ensembles and other bosonic systems such as microwave cavity QED.

\section{Hybrid quantum measurement scheme}

\begin{figure}[tb]
\includegraphics[width=1.0\hsize]{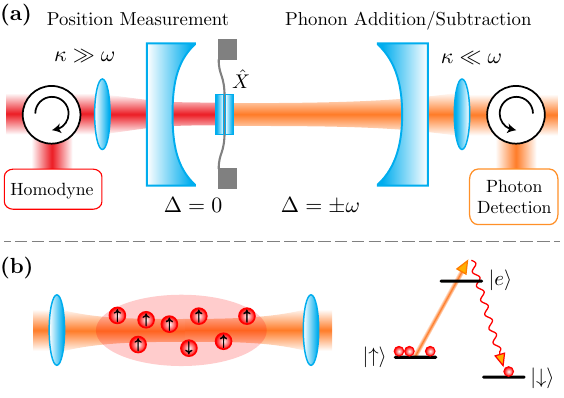}
\caption{(Colour online.)
Example bosonic systems that can implement our hybrid quantum measurement protocol.
(a)~An opto-mechanical system with a single mechanical oscillator (position quadrature $\hat{X}$) interacts via radiation pressure with two cavity modes. On the left is a short cavity for mechanical position measurements where the cavity decay rate $\kappa$ is much larger than the mechanical angular frequency $\omega$. A short cavity simultaneously allows for high finesse and large bandwidth allowing a rapid measurement of the mechanical position using a drive pulse with zero detuning ($\Delta = 0$) followed by an optical phase quadrature measurement with homodyne interferometry. On the right is a long cavity for interaction in the resolved-sideband regime ($\kappa \ll \omega$) for phonon addition or subtraction. These operations require detuning at the blue or red sidebands ($\Delta = \pm \omega$) and a single photon detection event heralds the addition or subtraction of a phonon, respectively.
(b)~Left: An atomic spin ensemble also provides an ideal experimental framework to implement this quantum-measurement-based state preparation scheme. QND quadrature measurements can be performed by polarization based homodyne detection and single quantum addition or subtraction are possible with single photon detection. Right: Atomic level diagram showing single quanta addition, i.e., transfer of a spin from the initial state $\ket{\uparrow}$ to $\ket{\downarrow}$ via an excited state $\ket{e}$ by detection of a scattered photon denoted by the wiggly red line.
}
\label{Fig:Scheme}
\end{figure}

An opto-mechanical setup to allow for both strong position measurements and single phonon addition or subtraction is shown in Fig.~\ref{Fig:Scheme}(a). The setup comprises a single mechanical oscillator that couples via radiation pressure to two independent optical cavity modes with quite different parameters. On the left side is a high-bandwidth cavity that allows for pulsed QND mechanical position measurements~\cite{Vanner2011}. This type of measurement requires operation in the regime where the cavity decay rate is much larger than the mechanical angular frequency, i.e., $\kappa \gg \omega$ (which is sometimes referred to as the `bad-cavity regime'). This requirement is to ensure that the mechanical position is negligibly changed, and hence unperturbed, by the opto-mechanical back-action noise during the pulsed interaction. After the pulse has reflected from the cavity the phase quadrature of the pulse is measured by homodyne detection in order to estimate the mechanical position. The action of such a measurement can be described, in the linearised regime, by the measurement operator
\begin{equation}\label{Eq:PosUps}
	\hat{\Upsilon}(\chi, \PL) = \pi^{-1/4} \exp [- {\textstyle\frac{1}{2}}(\PL - \chi \hat{X})^2 ] \, ,
\end{equation}
where $\PL$ is the measurement outcome, $\hat{X} = (\hat{b}+\hat{b}^\dagger)/\sqrt{2}$ is the mechanical position operator, $\chi$ quantifies the strength of the position measurement~\cite{ClassicalKickFootnote}, and $\hat{b}$ is the phonon annihilation operator. For a given initial mechanical state of motion $\rho_{\textrm{in}}$ the state following the measurement is given by $\rho_{\textrm{out}}(\PL) = [\hat{\Upsilon}(\chi, \PL) \circ \rho_{\textrm{in}}] / \text{Pr}(\PL)$, where $\text{Pr}(\PL) = \operatorname{Tr} [\hat{\Upsilon}(\chi, \PL) \circ \rho_{\textrm{in}}]$ is the probability distribution for the measurement outcome~\cite{Caves1987}, and the circle denotes action (i.e., $\hat{O} \circ \rho = \hat{O} \rho \hat{O}^\dag$). On the right side of the double cavity setup in Fig.~\ref{Fig:Scheme}(a) is a narrow-linewidth cavity that allows for phonon addition or subtraction via an optical drive on the blue or red sideband, respectively, followed by detection of a single photon spectrally selected on the cavity resonance, as was proposed in Ref.~\cite{VannerPRL2013}. Here, when a photon is detected with a frequency shifted down or up by $\omega$, due to energy conservation, a phonon must have been added or subtracted, respectively. This scheme requires operation in the resolved-sideband regime, i.e., $\kappa \ll \omega$, so that addition or subtraction can be individually performed, and, furthermore, the opto-mechanical coupling strength and the mechanical excitation must be weak in order to minimise multi-phonon excitations.  In this case the measurement operator for phonon subtraction is ${\textstyle \frac{\theta}{2}}\hat{b}$ and the measurement operator for phonon addition is $\varepsilon\hat{b}^\dagger$ where ${\textstyle \frac{\theta}{2}}$ is the effective opto-mechanical beam-splitter parameter, and $\varepsilon$ is the effective two-mode squeezing parameter.

These two measurement-based operations can also be realised in an atomic spin ensemble (cf.\@ Fig.~\ref{Fig:Scheme}(b)), which, for a large number of atoms, can be approximated as a bosonic mode. For this case, the analogue of the position measurement, Eq.~\eqref{Eq:PosUps}, can be achieved by measuring the polarisation rotation of an optical probe using a polarization based homodyne detection technique~\cite{Hammerer2010}. The analogue of single quantum addition or subtraction, i.e., single spin excitation, is achieved by detecting a scattered photon during a weak Raman-type excitation~\cite{Christensen2014, McConnell2015}. We would like to highlight at this point that our results and analysis are not restricted to describing particular experimental approaches to quadrature measurement or single quanta addition or subtraction, and can be readily applied to other approaches and bosonic systems that realize measurement operators of the same form.

We examine the non-classicality generated by permutations of these two measurement-based operations, where one is applied immediately after the other. In addition, we compare these cases to what can be generated with squeezing instead of a quadrature measurement. Specifically, we analyse the action of the following eight cases:
\begin{equation}\label{eq:operations}
	\begin{split}
		\hat{b}^{(\,,\dagger)} \hat{S}, \quad \textrm{and} \quad \hat{b}^{(\,,\dagger)} \hat{\Upsilon},\\
		\hat{S} \hat{b}^{(\,,\dagger)}, \quad \textrm{and} \quad \hat{\Upsilon} \hat{b}^{(\,,\dagger)},
	\end{split}
\end{equation}
where $\hat{S}(r) = \exp[{\textstyle\frac{1}{2}}r(\hat{b}^2 - \hat{b}^{\dagger 2})]$ is the single mode squeezing operator with squeezing parameter $r$, and the superscripts in brackets indicate that both addition or subtraction can be performed.

\section{Model and Figures of Merit}
\label{sec:modelEtc}

A particularly useful tool for understanding quantum states of a harmonic mode is the quasi-probability distribution.  The most common examples of quasi-probability distributions are the Glauber-Sudarshan \mbox{$P$-,} Wigner \mbox{$W$-,} and Husimi $Q$-functions~\cite{Barnett2002}.  The $P$-function, which given a density matrix $\rho$, is defined via
\begin{equation}
	\rho = \int d^2\beta P(\beta) \ket{\beta}\bra{\beta} .\label{eq:Pfunction}
\end{equation}
The $P$-, $W$-, and $Q$-functions correspond to normal, symmetric, and anti-normal ordering, respectively. Also note that the $W$-function's marginals are probability distributions. In fact, all three of these quasi-probability distributions are special cases of the generalised Cahill $R$-function (the $R$-function is similar to an $s$-parameterised Wigner function). Following Ref.~\cite{Lee1991}, we define an $R$-function from the $P$-function thus
\begin{equation}
\label{eq:Rdef}
    R_\tau(\beta) = \frac{1}{\pi \tau} \int \! d^2\beta' \exp(- |\beta - \beta'|^2 / \tau) P(\beta') .
\end{equation}
Substituting $\tau = 0$,~${\textstyle\frac{1}{2}}$,~or~$1$ yields the $P$-, $W$-, or $Q$-function respectively. More generally, $\tau$ can take on arbitrary values to describe distributions between the $P$-, $W$-, and $Q$-functions.

Non-classicality of a quantum state can be readily quantified once the $R$-function is known. A state may be deemed non-classical if the $P$-function does not exist. Additionally, a sufficient criterion for non-classicality is negativity in $W$-function. Of the $P$-, $W$-, and $Q$-distributions, only the $Q$-function both always exists and is always positive, thus constituting an acceptable probability distribution. To quantify non-classicality in our scheme we use two measures: (i) Wigner negativity (the total integrated negativity in the $W$-function) and (ii) non-classical depth (roughly, the minimum $\tau$ such that $R_\tau$ is an acceptable probability distribution). Both of these measures are defined more precisely below. It should also be highlighted that since the $W$-function always exists and completely characterizes a quantum state, it is a useful tool for visualising quantum states.

Before any operations are applied to the mechanical oscillator it is assumed to be in a thermal state with occupation $\bar{n}$. We may describe this with the density matrix  $\rho_{\bar{n}} = (1 + \bar{n})^{-1} [\bar{n} / (1 + \bar{n})]^{\hat{b}^\dag \hat{b}}$~\cite{Barnett2002}.  From Eqs.~\eqref{eq:Pfunction} and~\eqref{eq:Rdef} the $R$-function for a thermal state is
\begin{equation}
	R_{\tau, \bar{n}}(\beta) = [\pi (\tau + \bar{n})]^{-1} \exp[- |\beta|^2 / (\tau + \bar{n})] .
\end{equation}

By using commutation relations for $\hat{S}$ and $\hat{b}^{(, \dag)}$, all the squeezing cases in Eq.~(\ref{eq:operations}) may be written in the form
\begin{equation}\label{eq:squeezing form}
	{\textstyle \rho_{\hat{S}} = \sqrt{f_{\hat{b}^{(, \dag)}}} \hat{S}(r) (\nu \hat{b} + \mu \hat{b}^\dag) \circ \rho_{\bar{n}}} ,
\end{equation}
where $\sqrt{f_{\hat{b}^{(, \dag)}}} = {\textstyle\frac{\theta}{2}}$ or $\varepsilon$ depending upon whether we are subtracting or adding a phonon respectively.  The constants $\nu$ and $\mu$ for each squeezing case---$\hat{S} \hat{b}^{(, \dag)}$ or $\hat{b}^{(, \dag)} \hat{S}$---are listed in Appendix~\ref{sec:the constants}.  Similarly, by decomposing the measurement operator as $\hat{\Upsilon}(\chi, \PL) \circ \rho_{\bar{n}} = \sqrt{f_{\hat{\Upsilon}}({\PL})} \hat{S}(\xi) \hat{D}(\zeta \PL) \circ \rho_{\bar{m}}$~\cite{Vanner2011} where $\hat{D}(\delta) = \exp(\delta \hat{b}^\dag - \delta^* \hat{b})$ is the displacement operator and
\begin{gather}
	f_{\hat{\Upsilon}}(\PL) = (2 \pi \sigma_{\text{L}}^2)^{-1 / 2} \exp[- \PL^2 / (2 \sigma_{\text{L}}^2)] ,\\
	2 \sigma_{\text{L}}^2 = 1 + \chi^2 (1 + 2 \bar{n}) ,\\
	\xi = {\textstyle\frac{1}{4}} \log\{[\chi^2 + (1 + 2 \bar{n})] [\chi^2 + (1 + 2 \bar{n})^{-1}]\} ,\\
	\zeta = [\chi^2 + (1 + 2 \bar{n})^{-1}]^{-1} \chi e^{\xi} \text{, and}\\
	\bar{m} = {\textstyle\frac{1}{2}}\{\sqrt{[\chi^2 + (1 + 2 \bar{n})] [\chi^2 + (1 + 2 \bar{n})^{-1}]^{-1}} - 1\} ,
\end{gather}
all the quadrature measurement cases in (\ref{eq:operations}) may be written in the form
\begin{equation}\label{eq:measurement form}
	\begin{aligned}
		&\rho_{\hat{\Upsilon}}(\PL) \\
		&= {\textstyle \sqrt{f_{\hat{b}^{(, \dag)}} f_{\hat{\Upsilon}}(\PL)} \hat{S}(\xi) \hat{D}(\zeta \PL) (\nu \hat{b} + \mu \hat{b}^\dag + \lambda \PL) \circ \rho_{\bar{m}} } .
	\end{aligned}
\end{equation}
The constants $\nu$, $\mu$, and $\lambda$ for each measurement case---$\hat{\Upsilon} \hat{b}^{(, \dag)}$ or $\hat{b}^{(, \dag)} \hat{\Upsilon}$---are listed in Appendix~\ref{sec:the constants}.  Motivated by the fact that if $\PL = 0$ then Eqs.~\eqref{eq:squeezing form} and~\eqref{eq:measurement form} are equivalent, we identify $\xi$ as an \emph{effective} squeezing parameter and $\bar{m}$ as an \emph{effective} thermal occupation.  These forms for squeezing and measurement are particularly convenient, however the explicit expressions of the corresponding $R$-functions, which we denote $R_{\tau, \hat{S}}$ and $R_{\tau, \hat{\Upsilon}}$ respectively, are somewhat cumbersome and have therefore been relegated to Appendix~\ref{sec:general formula}.

In the above we have been careful to keep the $R$-functions $R_{\tau, \hat{S}}$ and $R_{\tau, \hat{\Upsilon}}$ non-normalised such that their probability distributions are $\text{Pr}_{\hat{S}} = \int d^2\beta R_{\tau, \hat{S}}(\beta)$ and $\text{Pr}_{\hat{\Upsilon}}(\PL) = \int d^2\beta R_{\tau, \hat{\Upsilon}}(\beta; \PL)$ respectively~\cite{Caves1987}.  Since phonon addition or subtraction is discrete and squeezing is deterministic, the heralding probability of the squeezing cases $p_{\hat{S}}$ is simply $\text{Pr}_{\hat{S}}$.  One easily finds
\begin{equation}
	p_{\hat{S}} = f_{\hat{b}^{(, \dag)}} [\nu^2 \bar{n} + \mu^2 (1 + \bar{n})] .
\end{equation}
On the other hand, the measurement outcome $\PL$ is a continuous variable, which means that demanding a particular outcome has a vanishing probability, and, instead, we must choose a window of outcomes $\PL \in (- w, w)$~\cite{fnFeedback}.  The heralding probability of the measurement cases $p_{\hat{\Upsilon}}$ is then the probability of obtaining an outcome within this window, i.e., $p_{\hat{\Upsilon}}(w) = \int_{- w}^{w} d\PL \text{Pr}_{\hat{\Upsilon}}(\PL)$.  After some computation,
\begin{equation}
	\begin{aligned}
		p_{\hat{\Upsilon}}(w)
			= f_{\hat{b}^{(, \dag)}} \Big\{& {\textstyle
		[\nu^2 \bar{m} + \mu^2 (1 + \bar{m}) + \lambda^2 \sigma_{\text{L}}^2] \text{erf}(w / \sqrt{2 \sigma_{\textrm{L}}^2})} \\
		&{\textstyle - 2 \lambda^2 \sigma_{\text{L}}^2 w f_{\hat{\Upsilon}}(w)
		} \Big\} ,
    \end{aligned}\label{eq:UpsilonNorm}
\end{equation}
where $\text{erf}$ is the error function~\cite{Abramowitz1972}.  The state to which such a probability corresponds is the statistical mixture of all states with $\PL \in (-w, w)$, and is therefore represented by the $R$-function
\begin{equation}\label{eq:mixture R-function}
	\bar{R}_{\tau, \hat{\Upsilon}}(\beta; w) = \int_{- w}^{w} d\PL R_{\tau, \hat{\Upsilon}}(\beta; \PL) .
\end{equation}
(See Appendix~\ref{sec:statistical formula} for the calculation of $\bar{R}_{\tau, \hat{\Upsilon}}$.)

We now define each of the figures of merit used here to quantify non-classicality.  The Wigner negativity $\delta$ of a state is the total integrated negative region of its normalised $W$-function.  Given a normalised $W$-function $W = R_{\tau = 1/2}$ this may be defined thus:
\begin{equation}
	\delta = {\frac{1}{2}}\left( \int d^2\beta |W(\beta)| - 1 \right) .\label{eq:WignerNegativity}
\end{equation}
The non-classical depth $\tau_\textrm{inf}$ of some state represented by the $R$-function $R_\tau$ is the infimum of the set of $\tau$s such that $R_\tau$ is an acceptable probability distribution.  Being an acceptable probability distribution may be defined quite abstractly, but in our case we require only to check the two most basic properties: that $R_\tau$ be integrable and non-negative.  Then, given an $R$-function $R_{\tau}$, the non-classical depth may be defined thus:
\begin{equation}
	\tau_{\textrm{inf}} = \operatorname{inf} \left\{ \tau : R_{\tau} \text{ is integrable and non-negative} \right\} .\label{eq:nonClassicalDepth}
\end{equation}
Finally, the fidelity with the general cat state $\kets{\text{cat}} = N_{\text{cat}}^{-1/2} [\kets{\beta_{\text{cat}}} + \exp(i \phi_{\text{cat}}) \kets{- \beta_{\text{cat}}}]$, where $N_{\text{cat}} = 2 [1 + \exp(- 2 |\beta_{\text{cat}}|^2) \cos \phi_{\text{cat}}]$, is $F = \bras{\text{cat}} \rho \kets{\text{cat}}$. The optimal fidelity is found by maximising the fidelity over $\beta_{\text{cat}}$ and $\phi_{\text{cat}}$. In our case it follows from symmetry that the optimal cat state has $\Re \beta_{\text{cat}} = 0$ and $\phi_{\text{cat}} = \pi$, and so we require only to optimise over $\Im \beta_{\text{cat}}$. In Ref.~\cite{Fiurasek2013} it is shown that the fidelity with a cat state is closely related to the $Q$-function.  Following the treatment therein, we may define our optimal fidelity of a state represented by the normalised $Q$-function $Q = R_{\tau = 1}$ thus:
\begin{equation}
\label{Eq:CatFidelity}
	\begin{aligned}
		F = \underset{\beta_{\text{cat}}}{\text{max}}
    \frac{\pi}{N_{\text{cat}}} \Big[& Q(\beta_{\text{cat}}) + Q(- \beta_{\text{cat}}) \\
    	&- 2 \Re \exp(- 2 |\beta_{\text{cat}}|^2) Q(\beta_{\text{cat}})|_{\beta_{\text{cat}}^* \mapsto - \beta_{\text{cat}}^*} \Big] ,
    \end{aligned}
\end{equation}
where it is assumed that $\Re \beta_{\text{cat}} = 0$~\cite{ComplexMapping}.

\section{Results and Discussion}
\label{sec:Results and Discussion}

\begin{figure*}[tb]%
	\begin{center}
        \includegraphics[width=\textwidth]{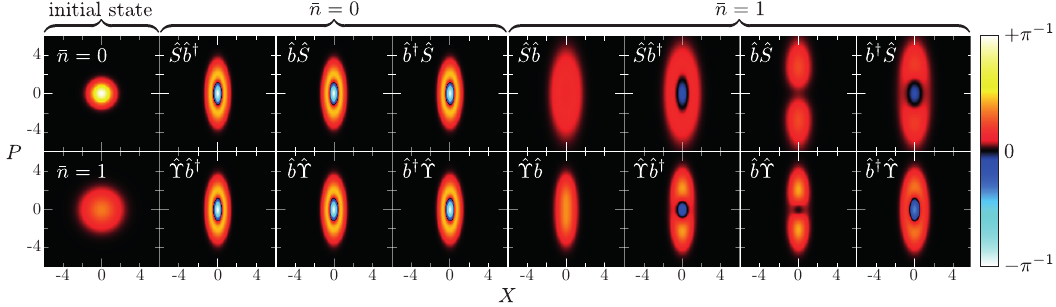}
		\caption{(Colour online.) Plots of example Wigner functions of states generated by our scheme. The first column contains the $W$-functions for initial thermal states with $\bar{n} = 0$ and $1$ as indicated. The other plots show the $W$-function after application of one of the eight cases in Eq.~\eqref{eq:operations}. Note that the cases $\hat{S} \hat{b}$ and $\hat{\Upsilon} \hat{b}$ are not included for $\bar{n} = 0$ because $\hat{b} \circ \rho_{\bar{n} = 0} = \hat{b} \kets{0} \bras{0} \hat{b}^\dag = 0$. The horizontal and vertical axes are the $X$ and $P$ quadratures, respectively, and the span of each is $[-6, 6]$ in every plot. For these plots, we have chosen: $r = 0.5$, $f_{\hat{b}^{(, \dag)}} = 10^{-2}$, and $w$ such that $p_{\hat{\Upsilon}}(w) = 10^{-4}$. The measurement strength $\chi$ is set by the relation $r = \xi$; one has $\chi \approx 1.31$ for $\bar{n} = 0$, and $\chi \approx 1.17$ for $\bar{n} = 1$. We note that all squeezing cases have a total heralding probability $p_{\hat{S}}$ of order $10^{-2}$, and all measurement cases require a window $w$ of order $10^{-2}$.}
		\label{fig:Wigner}
	\end{center}
\end{figure*}

Suppose for the moment that we may select $\PL = 0$.  Then the difference between the squeezing cases~\eqref{eq:squeezing form} and the quadrature measurement cases~\eqref{eq:measurement form} lies in the difference between the thermal occupation $\bar{n}$ and the effective thermal occupation $\bar{m}$, and the difference between the squeezing parameter $r$ and the effective squeezing parameter $\xi$.  Supposing also that $\bar{n} = 0$, the  quadrature measurement cases are then qualitatively identical to the squeezing cases and it is therefore pertinent to compare these two cases via the identification
\begin{equation}\label{eq:identification}
	\xi = r .
\end{equation}
Note, however, that for $\bar{n} = 0$ the cases $\hat{S} \hat{b}$ and $\hat{\Upsilon} \hat{b}$ have a heralding probability of zero, i.e., $\hat{b} \circ \rho_{\bar{n} = 0} = \hat{b} \kets{0} \bras{0} \hat{b}^\dag = 0$.

\begin{figure*}[tb]%
	\begin{center}
		\includegraphics[width=\textwidth]{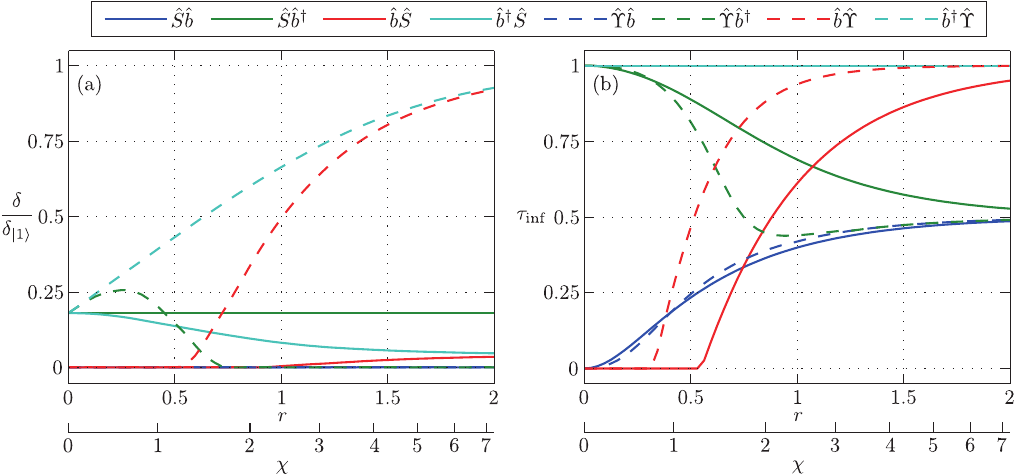}
		\caption{(Colour online.)  Plots of our two measures of non-classicality, viz.: (a)~Wigner negativity $\delta$ scaled by the Wigner negativity of a single quanta state $\delta_{\kets{1}}$; (b) non-classical depth $\tau_{\textrm{inf}}$.  We have chosen $\bar{n} = 1$ and otherwise all parameters are as in Fig.~\ref{fig:Wigner}.}
		\label{fig:WigNegTauPlots}
	\end{center}
\end{figure*}

In Fig.~\ref{fig:Wigner} we plot two example $W$-functions for each of the eight cases in Eq.~\eqref{eq:operations}: one for $\bar{n} = 0$ and one for $\bar{n} = 1$. (We will discuss behaviour with $\bar{n}$ more generally below.)  For these plots we have set $f_{\hat{b}^{(, \dag)}} = 10^{-2}$, and the quadrature measurement window $(-w, w)$ has been chosen such that $p_{\hat{\Upsilon}}(w) = 10^{-4}$. In the left-most column we have plotted the two initial states used: the ground state ($\bar{n}=0$), and a small thermal state ($\bar{n}=1$). The  six cases for $\bar{n}=0$ that have a non-zero heralding probability are equivalent and have Wigner negativity $\delta_{\kets{1}} = 2 e^{-1/2} - 1 \approx 0.2$. This value is equal to the Wigner negativity of a single quanta Fock state $\kets{1}$~\cite{Kenfack2004} and we henceforth scale Wigner negativity by $\delta_{\kets{1}}$ as an aid to intuition. The eight cases acting on the $\bar{n}=1$ thermal state show quite different phase-space features compared to the states prepared by acting on the ground state. Firstly, we note that the four cases that use $\hat{b}^\dagger$ show regions of significant Wigner negativity and resemble a cat state, whereas the four other cases which use $\hat{b}$ do not generate negativity for this thermal occupation. We also see that, now for finite thermal occupation, the cases $\hat{S}\hat{b}$ and $\hat{\Upsilon}\hat{b}$ have a finite heralding probability and generate states which are approximately Gaussian. Reversing the order of the operations for these two cases, i.e., performing $\hat{b} \hat{S}$ and $\hat{b} \hat{\Upsilon}$, generates a non-Gaussian distribution, which however shows no negativity owing to the initial thermal occupation. Note that $\hat{b} \hat{\Upsilon}$ in comparison to $\hat{b} \hat{S}$ has a deeper dip in the centre of phase-space, which becomes negative if the thermal occupation is reduced.

In Fig.~\ref{fig:WigNegTauPlots}(a) we plot the Wigner negativity (Eq.~\eqref{eq:WignerNegativity}) of the states generated by our eight cases as a function of the squeezing parameter and measurement strength for an initial thermal state with $\bar{n}=1$. It is evident that the operation $\hat{b}^\dagger \hat{\Upsilon}$ gives the deepest negativity. This is because performing a quadrature measurement first purifies the state before the negativity is generated by the phonon addition operation. If instead phonon subtraction is performed after the quadrature measurement, i.e., case $\hat{b} \hat{\Upsilon}$, then we see that the measurement strength must exceed a threshold for Wigner negativity to be generated. The value of this threshold is reduced by decreasing the initial thermal occupation. In the limiting case of large measurement strength both $\hat{b}^\dagger \hat{\Upsilon}$ and $\hat{b} \hat{\Upsilon}$ will converge and asymptote to $\delta_\textrm{\ket{1}}$. The case $\hat{\Upsilon}\hat{b}^\dagger$ shows a non-monotonic behaviour in the Wigner negativity owing to the trade-off between state purification, which dominates for small $\chi$, and the `filtering' of the position distribution as $\chi$ increases. The last quadrature measurement case, $\hat{\Upsilon}\hat{b}$, generates no Wigner negativity. For the squeezing cases, it is noted that the Wigner negativity of a quantum state is unchanged by a squeezing operation. Thus, the two cases, $\hat{S}\hat{b}^{(,\dagger)}$, are constant in $r$. On the other hand, performing $\hat{b}^{(,\dagger)} \hat{S}$ does show a dependence upon $r$ and the two cases converge in the limit of large $r$.

In Fig.~\ref{fig:WigNegTauPlots}(b) we plot the non-classical depth (Eq.~\eqref{eq:nonClassicalDepth}) using the same parameters as in Fig.~\ref{fig:WigNegTauPlots}(a). The non-classical depth quantitatively describes the robustness of a quantum state to added thermal noise~\cite{Lee1991} and shows different trends to Wigner negativity. Of the squeezing cases, the case $\hat{b}^\dagger\hat{S}$ generates a state with maximum non-classical depth $(\tau_\textrm{inf}=1)$, which is independent of $r$. The operation $\hat{b}\hat{S}$ approaches this maximum value for large squeezing and also shows a threshold behaviour similar to the state's Wigner negativity. Unlike Wigner negativity, however, the non-classical depth of a quantum state is affected by squeezing. Indeed, we see that the cases $\hat{S} \hat{b}^\dagger$ and $\hat{S} \hat{b}$ asymptote to $\tau_\textrm{inf}=1/2$ from the maximum $\tau_{\text{inf}} = 1$ and the minimum $\tau_{\text{inf}} = 0$ respectively. The cases using quadrature measurement are qualitatively similar to the cases using squeezing, but with the important exception that they generate similar non-classical depths for smaller values of the measurement strength in comparison to the (deterministic) squeezing parameter.

It is instructive to compare Fig.~\ref{fig:WigNegTauPlots}(a) with Fig.~\ref{fig:WigNegTauPlots}(b), for Wigner negativity and non-classical depth are complementary but not equivalent measures of non-classicality. One important property to note is that the Wigner negativity is non-zero when the non-classical depth is greater than $1/2$. It is also important to note that the asymptotic behaviour is very different for these two measures of non-classicality, and having a large non-classical depth does not guarantee a large Wigner negativity. A striking example is the case $\hat{b}^\dagger \hat{S}$ which gives the maximum non-classical depth yet has a comparatively small Wigner negativity. On the other hand, the case $\hat{b}^\dagger \hat{\Upsilon}$ has maximal non-classical depth and also gives the largest Wigner negativity of our eight cases.

\begin{figure}[tb]%
	\begin{center}
		\includegraphics[width=\columnwidth]{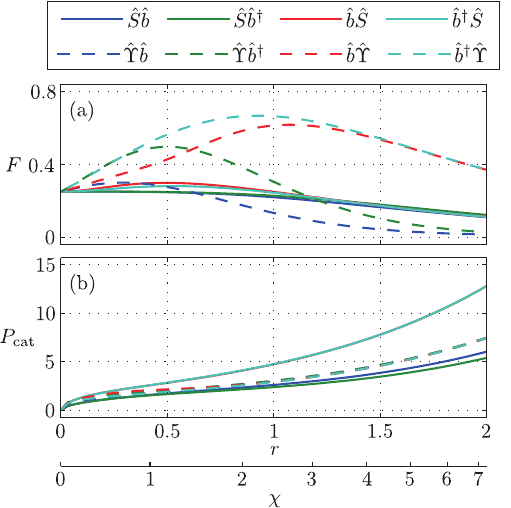}
		\caption{(Colour online.)  (a) Plot of the optimal fidelity with a cat state $F$, and (b) the corresponding cat-state separation $P_{\text{cat}} = (\beta_{\text{cat}} - \beta_{\text{cat}}^*) / \sqrt{2} i$.  All parameters are as in Fig.~\ref{fig:Wigner}.}
		\label{fig:extraFigure2}
	\end{center}
\end{figure}

In Fig.~\ref{fig:extraFigure2}(a) we plot the optimal fidelity with a cat state (Eq.~\eqref{Eq:CatFidelity}).  In order for such a fidelity to be meaningful, one must also verify that the amplitude of the associated cat state be significant; this is plotted in Fig.~\ref{fig:extraFigure2}(b) and it is evident that for the entire interesting range of $r$ or $\chi$ the amplitude is indeed significant.  As stated in Sec.~\ref{sec:modelEtc}, the particular cat state that yields this optimal fidelity has phase $\phi_{\text{cat}} = \pi$.  This is in fact the canonical cat state $\kets{\beta_{\text{cat}}} - \kets{- \beta_{\text{cat}}}$, and the fidelity with such a state is a known indicator of both Wigner negativity and non-Gaussianity~\cite{Fiurasek2013}.  Although fidelity with a cat state has limited utility for pure states, since for such states the fidelity is usually close to unity, it is quite insightful for the states with finite impurity considered here.  Optimal fidelity with a cat state also has one important qualitative difference to both Wigner negativity and non-classical depth: pronounced non-monotonicity.  As $r \rightarrow 0$ all cases converge to the same non-vanishing value, and as $r \rightarrow \infty$ the optimal fidelity with a cat state vanishes.  However, in between these limits the various cases separate and many attain a maximum for non-vanishing $r$.  This general behaviour is indicative of an extra competition with the fact that the overlap of the peaks with a cat state decreases as the squeezing or effective squeezing increases.  Overall, one sees that the squeezing and measurement cases do not behave qualitatively differently, but the measurement cases fare considerably better quantitatively.  We attribute this to the purification effected by measurement.  In harmony with both Wigner negativity and non-classical depth, it is easily seen that the wisest choice to create a state most similar to a cat state is $\hat{b}^\dag \hat{\Upsilon}$.

\begin{figure}[tb]%
	\begin{center}
		\includegraphics[width=\columnwidth]{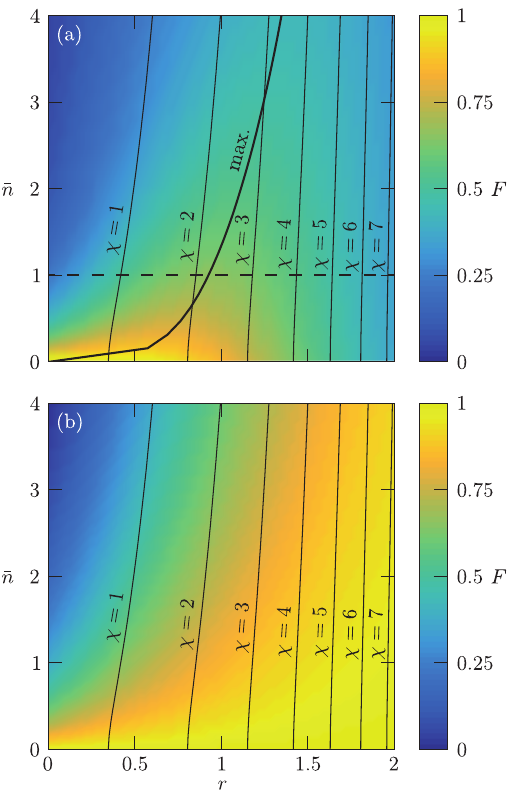}
		\caption{(Colour online.)  (a) Plot of the optimal fidelity with a cat state for the case $\hat{b}^\dag \hat{\Upsilon}$ vs.\@ $r$ and $\bar{n}$.  Lines of constant $\chi$ are included (recall that $\chi$ is determined via $\xi = r$).  The solid line labeled `max.' is the maximum in $r$ for given $\bar{n}$.  Fig.~\ref{fig:extraFigure2}(a) case $\hat{b}^\dag \hat{\Upsilon}$ corresponds to a slice along $\bar{n} = 1$ as indicated by the dashed line.  (b) Plot of the optimal fidelity with a squeezed single quanta state for case $\hat{b}^\dag \hat{\Upsilon}$ vs.\@ $r$ and $\bar{n}$ using Eq.~\eqref{eq:FS1}.}
		\label{fig:extraFigure2extra}
	\end{center}
\end{figure}

Having established that $\hat{b}^\dag \hat{\Upsilon}$ maximises all three of our figures of merit for all initial thermal occupations, it is accordant to consider how well this case performs in general, i.e., over the full range of both $r$ and $\bar{n}$: see Fig.~\ref{fig:extraFigure2extra}(a).  It is apparent that given an occupation $\bar{n}$ there is a particular $\chi$ for which the optimal fidelity with a cat state attains a maximum.

Examining the optimal fidelity with a cat state is useful to highlight the advantage of quadrature measurement over squeezing when the initial state is impure.  For example, for $\bar{n} = 1$ we can achieve $F \approx 0.67$ by using the operation $\hat{b}^\dagger \hat{\Upsilon}$, which may be compared with $F \approx 0.31$ for the operation $\hat{b} \hat{S}$.  It should be noted, however, that our scheme can in fact generate states that have a high fidelity with a squeezed single quanta state, $\hat{S}(s) \ket{1}$, even in the presence of significant initial thermal occupation.  For the case $\hat{b}^\dag \hat{\Upsilon}$ we need not use a window as the quadrature measurement is performed first and outcomes different to zero may be redressed by feedback.  For this operation, the optimal fidelity with $\hat{S}(s) \ket{1}$ is achieved for $s = \xi$ and one finds
\begin{equation}
	F = \frac{\cosh^2 \xi + 2 [\bar{m} / (1 + \bar{m})]^2 \sinh^2 \xi}{(1 + \bar{m}) \cosh^2 \xi + \bar{m} \sinh^2 \xi} .\label{eq:FS1}
\end{equation}
In the limit $\chi \rightarrow 0$ one has the fidelity of a thermal state with a single quanta state $\kets{1}$, i.e. $\lim_{\chi \rightarrow 0} F = 1 / (1 + \bar{n})$.  On the other hand, in the limit $\chi \rightarrow \infty$ the effective thermal occupation asymptotes to zero, $\bar{m} \rightarrow 0$, and hence $F \rightarrow 1$ regardless of the initial thermal occupation $\bar{n}$.  In Fig.~\ref{fig:extraFigure2extra}(b) we plot $F$ vs.\@ both $r$ and $\bar{n}$.  An impressively large area with high fidelity is apparent.

\section{Experimental Feasibility}

Our scheme can be implemented in a number of different quantum optical systems by combining current state-of-the-art techniques. For instance, the cavity-based spin ensemble experiment detailed in Ref.~\cite{McConnell2015} is very well suited to implementing our scheme by additionally performing spin squeezing measurements, which is a well established technique~\cite{Kuzmich2000, Appel2009, Sewell2013}. For an opto-mechanical realization there are many different physical systems that could be employed including, photonic crystal cavities and micro-toroids~\cite{Aspelmeyer2014}. We will describe another example approach here using a bulk-acoustic-wave (BAW) vibration in a high-reflectivity mirror forming part of a Fabry-Perot cavity and discuss a parameter set similar to that used in Refs.~\cite{Vanner2011, VannerPRL2013}. Such mechanical BAW modes have also been considered in Ref.~\cite{Xia2014} and experimentally studied in Ref.~\cite{Kuhn2011}. For the setup shown in Fig.~\ref{Fig:Scheme}(a) there are a number of requirements that need to be met simultaneously. We consider a mechanical resonator with a BAW resonance of $\omega/2\pi = 100$~MHz and an effective mass of 10~ng. A QND position measurement can be performed by using a small-mode-volume high-bandwidth cavity to allow for pulsed interactions. We consider a micro-Fabry-Perot design operating at 1064~nm with a cavity length of 532~nm. For a finesse of $1.4\times10^4$ this gives a cavity decay rate 100 times larger than $\omega$ thus allowing for rapid position measurements. For a pulsed interaction with $10^9$ photons this gives a mechanical position measurement strength $\chi \approx 1.0$. Phonon addition and subtraction can be performed with a second 7.5~mm cavity, which for a finesse of $10^4$ gives a cavity linewidth $100$ times smaller than $\omega$ thus allowing the optomechanical sidebands to be clearly separated for photon counting. We would also like to note that a QND position measurement can also be implemented with a two-toned drive on this longrt cavity~\cite{Aspelmeyer2014}, which offers an alternate route to implement our scheme using a single cavity. Importantly, the mechanical oscillator should be cyrogenically cooled to minimise thermal decoherence. With cryogenic cooling to $300$~mK, the mechanical oscillator will have a thermal occupation of $\nbar \approx 60$, and so some simple laser precooling~\cite{Aspelmeyer2014} can be employed by driving the red sideband of the longer cavity before implementing our protocol. At this temperature a modest mechanical quality factor of $10^4$ gives $\nbar_\text{bath} / Q < 10^{-2}$, thus the mechanical oscillator will undergo a very small amount of decoherence for several hundred mechanical periods. Thermal decoherence smooths the phase-space distribution of the quantum state prepared and reduces any non-classicality. We have quantified and discussed this effect in more detail in Appendix~\ref{sec:ThermalHeating}, which can be readily incorporated into the mathematical framework used here. It should also be noted that the precise linear measurements utilised here also provide a route to perform quantum state reconstruction~\cite{Vanner2014} by measurement of the mechanical quadrature marginals.

\section{Conclusions and Outlook}

Building on the success of non-classical state generation via squeezing and photon addition and subtraction in optics we propose and analyze the use of QND quadrature measurements in combination with single quanta addition and subtraction. This opens an avenue to generate non-classical states in massive bosonic systems, such as the motion of mechanical oscillators and spin-ensembles. Unlike optical fields, experiments in such systems face the challenge of initial thermal occupation, and it is a non-trivial task to identify which permutation of QND quadrature measurement and single photon addition or subtraction generates the strongest non-classicality. A key advantage of using measurement instead of squeezing is that it provides purification in addition to squeezing, thus offering more resilience to initial thermal occupation. Indeed, of the eight cases we have considered here, we find that the operation $\hat{b}^\dagger\hat{\Upsilon}$ provides the strongest Wigner negativity, maximises the non-classical depth, and attains the largest optimal fidelity with a cat state. Our scheme can be immediately applied in atomic spin ensemble and optomechanics experiments where it can be useful for the development of quantum technologies, e.g., quantum sensing, and for probing fundamental physics by empirically studying decoherence mechanisms. Other further work in this direction could include a treatment of multiple operations, instead of a sequence of two, during the open-system dynamics using a quantum trajectory approach.

\section*{Acknowledgments}
We would like to acknowledge useful discussion with G.~J.~Milburn, I.~Pikovski, E.~S.~Polzik, P.~Rabl, J.~Twamley, and K.~Xia. This work was supported by an Australian Research Council Discovery Project (DP140101638), the OPSOQI project 316607 of the WWTF, the CoQuS doctoral programme DK CoQuS W 1210, the START grant Y 591-N16, and the UK Engineering and Physical Sciences Research Council (EPSRC EP/K034480/1). T.J.M. acknowledges the kind hospitality provided by The University of Queensland, the University of Oxford, and Imperial College London. M.R.V. acknowledges the kind hospitality provided by the Korean Institute of Advanced Study, the Technical University of Vienna, and Imperial College London. T.J.M. and M.R.V. jointly acknowledge `the Turf', whose refreshments were a \textit{sine qua non}.

\appendix

\section{List of the constants $\nu$, $\mu$, and $\lambda$ for each case}\label{sec:the constants}

Here we list the expressions for $\nu$, $\mu$, and $\lambda$ as introduced in Eqs.~\eqref{eq:squeezing form} and~\eqref{eq:measurement form} that correspond to the eight cases in Eq.~\eqref{eq:operations}.  For $\hat{S} \hat{b}$
\begin{equation}
    \nu = 1 \text{, } \mu = 0 \text{, and } \lambda = 0 ;
\end{equation}
for $\hat{S} \hat{b}^\dag$
\begin{equation}
    \nu = 0 \text{, } \mu = 1 \text{, and } \lambda = 0 ;
\end{equation}
for $\hat{b} \hat{S}$
\begin{equation}
    \nu = \cosh r \text{, } \mu = - \sinh r \text{, and } \lambda = 0 ;
\end{equation}
for $\hat{b}^\dag \hat{S}$
\begin{equation}
    \nu = - \sinh r \text{, } \mu = \cosh r \text{, and } \lambda = 0 ;
\end{equation}
for $\hat{\Upsilon} \hat{b}$
\begin{gather}
	\begin{gathered}
		\nu = \cosh \xi + \frac{\chi^2}{2} e^{- \xi} ,\\
		\mu = -\sinh \xi + \frac{\chi^2}{2} e^{- \xi} \text{, and}\\
		\lambda = - \frac{\chi}{\sqrt{2}}
		+ (1 + \chi^2) \zeta e^{- \xi} ;
	\end{gathered}
\end{gather}
for $\hat{\Upsilon} \hat{b}^\dag$
\begin{gather}
	\begin{gathered}
		\nu = -\sinh \xi - \frac{\chi^2}{2} e^{- \xi} ,\\
		\mu = \cosh \xi - \frac{\chi^2}{2} e^{- \xi} \text{, and}\\
		\lambda = \frac{\chi}{\sqrt{2}} + (1 - \chi^2) \zeta e^{- \xi} ;
    \end{gathered}
\end{gather}
for $\hat{b} \hat{\Upsilon}$
\begin{equation}
	\nu =  \cosh \xi \text{, } \mu = - \sinh \xi \text{, and } \lambda = \zeta e^{- \xi} ;
\end{equation}
for $\hat{b}^\dag \hat{\Upsilon}$
\begin{equation}
	\nu = - \sinh \xi \text{, } \mu = \cosh \xi \text{, and } \lambda = \zeta e^{- \xi} .
\end{equation}

\section{Calculation of the $R$-function representing $\hat{S}(\xi) \hat{D}(\zeta \PL) (\nu \hat{b} + \mu \hat{b}^\dag + \lambda \PL) \circ \rho_{\bar{m}}$}\label{sec:general formula}

As the form $\hat{S}(\xi) \hat{D}(\zeta \PL) (\nu \hat{b} + \mu \hat{b}^\dag + \lambda \PL) \circ \rho_{\bar{m}}$ suggests, we calculate its $R$-function representation, which we denote $R_{\tau, \xi, \zeta, \bar{m}, \nu, \mu, \lambda}$, in three stages: (i) we calculate the effect of $\nu \hat{b} + \mu \hat{b}^\dag + \lambda \PL$; (ii) the effect of $\hat{S}$; (iii) the effect of $\hat{D}$.  We save the displacement operator for last as its effect is trivial and only important for the measurement cases.  In order to do the calculation in this order, however, we must reverse the order of squeezing and displacement: $\hat{S}(\xi) \hat{D}(\zeta \PL) = \hat{D}(\zeta_\xi \PL) \hat{S}(\xi)$ where $\zeta_\xi = \zeta \cosh \xi - \zeta \sinh \xi$.

(i) The effect of $\nu \hat{b} + \mu \hat{b}^\dag + \lambda \PL$ is most readily found by focusing on the $P$-function.  This is because the $P$-function is an integral expansion in coherent states and the action of any polynomial in $\hat{b}$ and $\hat{b}^{\dag}$ on a coherent state may be rewritten as a differential operator on the same, thus affording a derivation of the resultant $P$-function by means of integration by parts. It is convenient for this purpose to introduce the so-called Bargmann coherent state $||\beta\rangle = \exp(\beta \hat{b}^\dag) \kets{0}$, in terms of which the usual coherent state is $\kets{\beta} = \exp(|\beta|^2 / 2) ||\beta\rangle$. The defining equation for a $P$-function given a density matrix $\rho$ then becomes
\begin{equation}
    \rho = \int d^2\beta \exp(- |\beta|^2) P(\beta) ||\beta\rangle\langle\beta|| .
\end{equation}
The utility of the Bargmann coherent state is due to the fact that it satisfies the following three relations:
\begin{equation}
    \hat{b} ||\beta\rangle = \beta ||\beta\rangle \text{, } \hat{b}^\dag ||\beta\rangle = \frac{\partial}{\partial \beta} ||\beta\rangle \text{, and } \frac{\partial}{\partial \beta^*} ||\beta\rangle = 0 .
\end{equation}
The action of $(\nu \hat{b} + \mu \hat{b}^\dag + \lambda \PL)$ on $\rho_{\bar{m}}$ results in nine terms.  As an example, let us calculate the $P$-function that represents $\hat{b} \rho_{\bar{m}} \hat{b}$ (recall that the $P$-function representing a thermal state $\rho_{\bar{m}}$ is $P_{\bar{m}}(\beta) = (\pi \bar{m})^{-1} \exp(- |\beta|^2 / \bar{m})$):
\begin{align}
    \hat{b} \rho_{\bar{m}} \hat{b} &= \int d^2\beta \exp(- |\beta|^2) P_{\bar{m}}(\beta) \hat{b} ||\beta\rangle \langle\beta|| \hat{b} \\
    &= \int d^2\beta \exp(- |\beta|^2) P_{\bar{m}}(\beta) \beta ||\beta\rangle \left[ \frac{\partial}{\partial \beta^*} \langle\beta|| \right] \\
    &= - \int d^2\beta \left[ \frac{\partial}{\partial \beta^*} \exp(- |\beta|^2) P_{\bar{m}}(\beta) \beta ||\beta\rangle \right] \langle\beta|| \\
    &= \int d^2\beta \frac{1 + \bar{m}}{\bar{m}} \beta \exp(- |\beta|^2) P_{\bar{m}}(\beta) \beta ||\beta\rangle \langle\beta|| \\
    &= \int d^2\beta \frac{1 + \bar{m}}{\bar{m}} \beta^2 P_{\bar{m}}(\beta) \kets{\beta} \bras{\beta} ,
\end{align}
whence the $P$-function that represents $\hat{b} \rho_{\bar{m}} \hat{b}$ is $[(1 + \bar{m}) / \bar{m}] \beta^2 P_{\bar{m}}(\beta)$.  Altogether, the final result for the $P$-function that represents $(\nu \hat{b} + \mu \hat{b}^\dag + \lambda \PL) \circ \rho_{\bar{m}}$ is
\begin{widetext}
\begin{equation}
    \begin{aligned}
        P_{\bar{m}, \nu, \mu, \lambda}(\beta; \PL) = &\left[ \nu^2 |\beta|^2 + \mu^2 (1 + \bar{m}) \frac{(1 + \bar{m}) |\beta|^2 - \bar{m}}{\bar{m}^2} + \lambda^2 \PL^2 \right.\\
        &\quad\left.+ \nu \mu \frac{1 + \bar{m}}{\bar{m}} (\beta^2 + \beta^{* 2}) + \nu \lambda \PL (\beta + \beta^*) + \mu \lambda \PL \frac{1 + \bar{m}}{\bar{m}} (\beta + \beta^*) \right] P_{\bar{m}}(\beta) .
    \end{aligned}
\end{equation}
\end{widetext}
Note that the normalisation is not unity but rather $\nu^2 \bar{m} + \mu^2 (1 + \bar{m}) + \lambda^2 \PL^2$.

(ii) The effect of $\hat{S}$ is simplest when considering the $W$-function, for which it simply rescales a certain pair of axes; in our case $\Re \beta$ and $\Im \beta$ by $e^{\xi}$ and $e^{- \xi}$ respectively.  Let us call $\Re \beta = x$ and $\Im \beta = y$ and rewrite $P_{\bar{m}, \nu, \mu, \lambda}$ in the form~\cite{HermiteFootnote}
\begin{equation}
    \begin{aligned}
    &P_{\bar{m}, \nu, \mu, \lambda}(\beta; \PL) \\
    &\quad= \sum_{n, m = 0}^{\infty} p_{n, m}(\PL) \frac{\partial^n}{\partial x^n} \frac{\partial^m}{\partial y^m} g_{0, \sigma_x}(x) g_{0, \sigma_y}(y)
    \end{aligned}
\end{equation}
where $p_{n, m}$ are constants and $g_{\mu, \sigma}$ is a Gaussian with mean $\mu$ and variance $\sigma^2$;
\begin{equation}
    g_{\mu, \sigma}(x) = (2 \pi \sigma^2)^{-1/2} \exp[- (x - \mu)^2 / (2 \sigma^2)] .
\end{equation}
In our case, $2 \sigma_x^2 = 2 \sigma_y^2 = \bar{m}$ and the only non-vanishing constants $p_{n, m}$ are the following:
\begin{gather}
	\begin{gathered}
    	p_{0, 0}(\PL) = \nu^2 \bar{m} + \mu^2 (1 + \bar{m}) + \lambda^2 \PL^2 ;\\
    	p_{1, 0}(\PL) = - \nu \lambda \PL \bar{m} - \mu \lambda \PL (1 + \bar{m}) ;\\
    	p_{2, 0}(\PL) = {\textstyle\frac{1}{4}} \nu^2 \bar{m}^2 + {\textstyle\frac{1}{4}} \mu^2 (1 + \bar{m})^2 + {\textstyle\frac{1}{2}} \nu \mu \bar{m} (1 + \bar{m}) ;\\
    	p_{0, 2}(\PL) = {\textstyle\frac{1}{4}} \nu^2 \bar{m}^2 + {\textstyle\frac{1}{4}} \mu^2 (1 + \bar{m})^2 - {\textstyle\frac{1}{2}} \nu \mu \bar{m} (1 + \bar{m}) .
    \end{gathered}
\end{gather}
From Eq.~\eqref{eq:Rdef}, the $W$-function that corresponds to the $P$-function $P_{\bar{m}, \nu, \mu, \lambda}$ is obtained via a double convolution with $g_{0, \sqrt{1 / 2}} g_{0, \sqrt{1 / 2}}$.  Using the theorem presented in Ref.~\cite{Fang1994}, this yields
\begin{equation}
    \begin{aligned}
        &W_{\bar{m}, \nu, \mu, \lambda}(\beta; \PL) \\
        &\quad= \sum_{n, m = 0}^{\infty} w_{n, m}(\PL) \frac{\partial^n}{\partial x^n} \frac{\partial^m}{\partial y^m} g_{0, \sigma_x^W}(x) g_{0, \sigma_y^W}(y)
    \end{aligned}
\end{equation}
where $w_{n, m} = p_{n, m}$ and $2 (\sigma_x^W)^2 = 2 (\sigma_y^W)^2 = \frac{1}{2} + \bar{m}$.  The effect of $\hat{S}$ may then be written $W_{\xi, \bar{m}, \nu, \mu, \lambda}(\beta; \PL) = W_{\bar{m}, \nu, \mu, \lambda}(\beta; \PL)|_{x \mapsto x e^{\xi}, y \mapsto y e^{- \xi}}$:
\begin{equation}
    \begin{aligned}
        &W_{\xi, \bar{m}, \nu, \mu, \lambda}(\beta; \PL) \\
        &\quad= \sum_{n, m = 0}^{\infty} w_{n, m}^\xi(\PL) \frac{\partial^n}{\partial x^n} \frac{\partial^m}{\partial y^m} g_{0, \sigma_x^\xi}(x) g_{0, \sigma_y^\xi}(y)
    \end{aligned}
\end{equation}
where $w_{n, m}^\xi = e^{- (n - m) \xi} w_{n, m}$, $2 (\sigma_x^\xi)^2 = (\frac{1}{2} + \bar{m}) e^{- 2 \xi}$, and $2 (\sigma_y^\xi)^2 = (\frac{1}{2} + \bar{m}) e^{2 \xi}$.

(iii) The effect of $\hat{D}$ regarding any quasi-probability distribution is simply to shift the origin by $\zeta_\xi \PL$, and the $R$-funtion that represents $W_{\xi, \bar{m}, \nu, \mu, \lambda}$ is obtained via a double convolution with $g_{0, \sqrt{(\tau - \frac{1}{2}) / 2}} g_{0, \sqrt{(\tau - \frac{1}{2}) / 2}}$.  Thus, again using the theorem presented in Ref.~\cite{Fang1994}, the $R$-function that represents $\hat{D}(\zeta_\xi \PL) \hat{S}(\xi) (\nu \hat{b} + \mu \hat{b}^\dag + \lambda \PL) \circ \rho_{\bar{m}}$ is
\begin{equation}\label{eq:fullR}
    \begin{aligned}
        &R_{\tau, \xi, \zeta, \bar{m}, \nu, \mu, \lambda}(\beta; \PL) \\
        &\quad= \sum_{n, m = 0}^{\infty} r_{n, m}(\PL) \frac{\partial^n}{\partial x^n} \frac{\partial^m}{\partial y^m} g_{\zeta_\xi \PL, \sigma_x^\tau}(x) g_{0, \sigma_y^\tau}(y) ,
    \end{aligned}
\end{equation}
where $r_{n, m} = w_{n, m}^\xi$ and
\begin{gather}
    2 (\sigma_x^\tau)^2 = {\textstyle (\frac{1}{2} + \bar{m}) e^{- 2 \xi} + \tau - \frac{1}{2}} \text{ and}\\
    2 (\sigma_y^\tau)^2 = {\textstyle (\frac{1}{2} + \bar{m}) e^{2 \xi} + \tau - \frac{1}{2}} .
\end{gather}

\section{Calculation of the $R$-function representing $\int_{-w}^{w} d\PL [\sqrt{f_{\hat{\Upsilon}}(\PL)}\hat{S}(\xi) \hat{D}(\zeta \PL) (\nu \hat{b} + \mu \hat{b}^\dag + \lambda \PL) \circ \rho_{\bar{m}}]$}\label{sec:statistical formula}

In Sec.~\ref{sec:general formula} is presented the $R$-function representing $\hat{S}(\xi) \hat{D}(\zeta \PL) (\nu \hat{b} + \mu \hat{b}^\dag + \lambda \PL) \circ \rho_{\bar{m}}]$, namely, $R_{\tau, \xi, \zeta, \bar{m}, \nu, \mu, \lambda}$.  Using this result, the $R$-function representing $\int_{-w}^{w} d\PL [\sqrt{f_{\hat{\Upsilon}}(\PL)}\hat{S}(\xi) \hat{D}(\zeta \PL) (\nu \hat{b} + \mu \hat{b}^\dag + \lambda \PL) \circ \rho_{\bar{m}}]$ may be written
\begin{equation}
    \begin{aligned}
        &\bar{R}_{\tau, \xi, \zeta, \bar{m}, \nu, \mu, \lambda}(\beta; w) \\
        &\quad= \int_{- w}^w d\PL f_{\hat{\Upsilon}}(\PL) R_{\tau, \xi, \zeta, \bar{m}, \nu, \mu, \lambda}(\beta; \PL) .
    \end{aligned}
\end{equation}
The following evaluation of this integral uses the same formalism as in Sec.~\ref{sec:general formula}, only since now the integral is over a sub-domain of $\mathbb{R}$ boundary terms appear.  Suppose we may expand thus
\begin{equation}
	f_{\hat{\Upsilon}}(\PL) r_{n, m}(\PL) = \sum_{l = 0}^{\infty} r_{n, m, l} \frac{\partial^l}{\partial \PL^l} g_{0, \sigma_{\textrm{L}}}(\PL)
\end{equation}
where $r_{n, m, l}$ are constants.  In our case, the only non-vanishing constants $r_{n, m, l}$ are the following:
\begin{gather}
	\begin{gathered}
    	r_{0, 0, 0} = \nu^2 \bar{m} + \mu^2 (1 + \bar{m}) + \lambda^2 \sigma_{\text{L}}^2 ;\\
    	r_{0, 0, 2} = \lambda^2 \sigma_{\text{L}}^4 ;\\
    	r_{1, 0, 1} = e^{- \xi} [\nu \lambda \bar{m} + \mu \lambda (1 + \bar{m})] \sigma_{\text{L}}^2  ;\\
    	r_{2, 0, 0} = e^{- 2 \xi} [{\textstyle\frac{1}{4}} \nu^2 \bar{m}^2 + {\textstyle\frac{1}{4}} \mu^2 (1 + \bar{m})^2 + {\textstyle\frac{1}{2}} \nu \mu \bar{m} (1 + \bar{m})] ;\\
    	r_{0, 2, 0} = e^{2 \xi} [{\textstyle\frac{1}{4}} \nu^2 \bar{m}^2 + {\textstyle\frac{1}{4}} \mu^2 (1 + \bar{m})^2 - {\textstyle\frac{1}{2}} \nu \mu \bar{m} (1 + \bar{m})] .
    \end{gathered}
\end{gather}
Then the expression for $\bar{R}_{\tau, \xi, \zeta, \bar{m}, \nu, \mu, \lambda}$ becomes
\begin{widetext}
\begin{equation}
	\bar{R}_{\tau, \xi, \zeta, \bar{m}, \nu, \mu, \lambda}(\beta; w) = \sum_{n, m, l = 0}^{\infty} r_{n, m, l} \frac{\partial^n}{\partial x^n} \frac{\partial^m}{\partial y^m} \left[
		\int_{-w}^{w} d\PL \frac{\partial^l g_{0, \sigma_{\text{L}}}(\PL)}{\partial \PL^l} g_{\zeta_\xi \PL, \sigma_x}(x)
		\right] g_{0, \sigma_y}(y) .
\end{equation}
\end{widetext}
Let's focus on the integral in brackets:
\begin{equation}
	I = \int_{-w}^{w} d\PL \frac{\partial^l g_{0, \sigma_{\text{L}}}(\PL)}{\partial \PL^l} g_{\zeta_\xi \PL, \sigma_x}(x) .
\end{equation}
Performing integration by parts $l$ times yields
\begin{equation}
	\begin{aligned}
		I =& \left[
		\sum_{k = 0}^{l - 1} (-1)^k \frac{\partial^{l - 1 - k} g_{0, \sigma_{\text{L}}}(\PL)}{\partial\PL^{l - 1 - k}} \frac{\partial^{k} g_{\zeta_\xi \PL, \sigma_x}(x)}{\partial \PL^{k}}
		\right]_{\PL = -w}^w \\
        &+ (-1)^l \int_{-w}^w d\PL g_{0, \sigma_{\text{L}}}(\PL) \frac{\partial^l g_{\zeta_\xi \PL, \sigma_x}(x)}{\partial\PL^l} .
    \end{aligned}
\end{equation}
The last term here may be computed by, firstly, exploiting the peculiar fact that $(\partial^k/\partial \PL^k) g_{\zeta_\xi \PL, \sigma_x}(x) = (- \zeta_\xi)^k (\partial^k/\partial x^k) g_{\zeta_\xi \PL, \sigma_x}(x)$, whereby one may swap the order of integration and differentiation, and then, secondly, applying 2.33.1 in Ref.~\cite{Gradshteyn2007}:
\begin{equation}
	\begin{aligned}
    	&(-1)^l \int_{-w}^w d\PL g_{0, \sigma_{\text{L}}}(\PL) \frac{\partial ^l g_{\zeta_\xi \PL, \sigma_x}(x)}{\partial \PL^l} \\
        &\quad = - (\zeta_\xi)^l \frac{\partial^l}{\partial x^l} G_{\varsigma^2 \zeta_\xi \PL, \varsigma \sigma_x}(x) g_{0, \varsigma \zeta_\xi \sigma_{\text{L}}}(x)
   	\end{aligned}
\end{equation}
where $\varsigma^2 = 1 + (\sigma_x / \zeta_\xi \sigma_{\text{L}})^2$ and $G_{\mu, \sigma}$ is the primitive of $g_{\mu, \sigma}$;
\begin{equation}
	G_{\mu, \sigma}(x) = \int dx g_{\mu, \sigma}(x) = {\textstyle\frac{1}{2}} \operatorname{erf}[(x - \mu)/\sqrt{2 \sigma^2}] .
\end{equation}
Thus, altogether we have
\begin{widetext}
\begin{equation}
	\begin{aligned}
		\bar{R}_{\tau, \xi, \zeta, \bar{m}, \nu, \mu, \lambda}(\beta; w) = \sum_{n, m, l = 0}^{\infty} r_{n, m, l} \frac{\partial^n}{\partial x^n} \frac{\partial^m}{\partial y^m} &\left[
			\sum_{k = 0}^{l - 1} (\zeta_\xi)^k \frac{\partial^{l - 1 - k} g_{0, \sigma_{\text{L}}}(\PL)}{\partial \PL^{l - 1 - k}} \frac{\partial^k g_{\zeta_\xi \PL, \sigma_x^\tau}(x)}{\partial x^k} \right.\\
            &\quad- \left. (\zeta_\xi)^l \frac{\partial^l}{\partial x^l} G_{\varsigma^2 \zeta_\xi \PL, \varsigma \sigma_x^\tau}(x) g_{0, \varsigma \zeta_\xi \sigma_{\text{L}}}(x) \right]_{\PL = - w}^w g_{0, \sigma_y^\tau}(y) .
	\end{aligned}
    \label{eq:bigFullR}
\end{equation}
\end{widetext}

\section{On the effect of thermal heating}\label{sec:ThermalHeating}

During our protocol, coupling of the system of interest to its surrounding thermal bath will cause an inevitable amount of decoherence. This has the effect of smoothing the phase-space distribution and reducing any negativity or non-classicality. In this appendix we discuss how this effect can be quantified using our framework.

Of the operations in our scheme, addition or subtraction, $\hat{b}^{(, \dag)}$, takes by far the longest time to perform. As described in Ref.~\cite{VannerPRL2013} this operation is achieved in the resolved sideband regime and requires an interaction time of many mechanical periods (of order 100). By contrast, a quadrature measurement, $\hat{\Upsilon}$, is performed over a timescale which is much shorter than a mechanical period. Thus, the principal contribution of thermal decoherence will occur during the action of $\hat{b}^{(, \dag)}$ and we neglect decoherence during $\hat{\Upsilon}$.

The mechanical rethermalisation rate is $\nbar_\text{bath} \gamma$ ($\nbar_\text{bath}$: thermal occupation of the bath; $\gamma$: mechanical damping rate) and thus a useful dimensionless constant is $\nbar_\text{bath} / Q$, which approximates the decoherence per mechanical period ($Q$: mechanical quality factor). We may therefore quantify the thermal decoherence during the action of $\hat{b}^{(, \dag)}$ as $\tau_{\text{th}} \approx N_T \times \nbar_{\text{bath}} / Q$---the number of thermal quanta `added' during $\hat{b}^{(, \dag)}$, where $N_T$ is the interaction time in mechanical periods and is of order 100 as described above.  Clearly, we require $\tau_{\text{th}} \ll 1$ in order for our protocol to generate non-classicality.  We have discussed in the main text that for modest experimental parameters one may achieve $\nbar_{\text{bath}} / Q < 10^{-2}$, for which this condition is satisfied.

The formalism we employ readily affords a simple model of thermal decoherence.  Since the decoherence due to thermal heating is assumed to be small we may treat it as a perturbation to the ideal evolution that effects $\hat{b}^{(, \dag)}$.  Therefore, the evolution may be separated and the full dynamics approximated by first applying $\hat{b}^{(, \dag)}$ and then performing partial rethermalisation by `adding' $\tau_{\text{th}}$ thermal quanta.  The operation that describes adding $\tau_{\text{th}}$ thermal quanta to some state $\rho$ is~\cite{Musslimani1995}
\begin{equation}
	\int \frac{d^2\beta}{\pi \tau_{\text{th}}} \exp(- |\beta|^2 / \tau_{\text{th}}) \hat{D}(\beta) \rho \hat{D}^\dag(\beta) .
\end{equation}
(One may derive this by adiabatically eliminating the bath, treating the decoherence as a perturbation, and finally assuming that the thermal occupation of the bath is much greater than unity, which allows the bath to be treated classically).  For the $R$-function that represents $\rho$, this becomes a convolution with a Gaussian of variance $\tau_{\text{th}}/ 2$.  Hence,
\begin{equation}
	R_{\tau} \underset{\substack{\text{thermal}\\\text{heating}}}{\mapsto} R_{\tau + \tau_{\text{th}}} .
\end{equation}
(This observation is in fact the motivation for interpreting the non-classical depth as the average number of thermal quanta required to eliminate negativity.)  Thus, the formalism used here with the $R$-function allows one to readily incorporate thermal heating into expressions such as Eqs.~\eqref{eq:fullR} and~\eqref{eq:bigFullR}.  Considering, for example, the case $\hat{b}^\dag \hat{\Upsilon}$, the effect of a non-vanishing $\tau_{\text{th}}$ is simply to smoothen the phase-space distribution, and reduce the non-classical depth by $\tau_{\text{th}}$.



\begin{thebibliography}{99}
\bibitem{fnFeedback} One might think that feedback could be used to displace the mechanical state back to the origin and thereby avoid the use of a window.  This works if the quadrature measurement is performed before addition or subtraction---cases $\hat{b}^{(,\dag)} \hat{\Upsilon}$---but if addition or subtraction comes first then the resulting state would be asymmetric and such feedback cannot be used.
\bibitem{ComplexMapping} Note that the mapping $\beta_{\text{cat}}^* \mapsto - \beta_{\text{cat}}^*$ is equivalent to $(\Re \beta, \Im \beta) \mapsto (i \Im \beta, - i \Re \beta)$.  Also, we use the convention that $\Re$ yields the real part and $\Im$ the imaginary.
\bibitem{HermiteFootnote} For numerical evaluation, the following relation is helpful:
\begin{equation}
    \frac{d^n}{dx^n} g_{\mu, \sigma}(x) = (-1 / \sqrt{2 \sigma^2})^n H_n[(x - \mu) / \sqrt{2 \sigma^2}] g_{\mu, \sigma}(x) \nonumber
\end{equation}
where $H_n$ is the physicists' Hermite polynomial of order $n$.
\end{thebibliography}
\end{document}